\documentclass[12pt]{article}

\newcommand{\m}{\mathrm}
\newcommand{\be}{\begin{equation}}
\newcommand{\ee}{\end{equation}}
\newcommand{\ba}{\begin{eqnarray}}
\newcommand{\ea}{\end{eqnarray}}

\usepackage{graphicx}
\usepackage{amssymb}
\usepackage{amsmath}
\usepackage[T1]{fontenc} 
\usepackage[ansinew]{inputenc} 
\usepackage[nosort]{cite}
\newcommand{\inbar}{\vrule height1.57ex width.4pt depth0pt}
\newcommand{\SW}{\relax{\hbox{$\ \inbar\kern-.285em{\rm S}$}}}

\begin{document}
\thispagestyle{empty}
\begin{center}

\null \vskip-1truecm \vskip2truecm

{\Large{\bf \textsf{Why is Black Hole Entropy Affected by Rotation?}}}

{\large{\bf \textsf{}}}

{\large{\bf \textsf{}}}

\vskip1truecm

{\large \textsf{Brett McInnes}}

\vskip1truecm

\textsf{\\  National
  University of Singapore}

\textsf{email: matmcinn@nus.edu.sg}\\

\end{center}
\vskip1truecm \centerline{\textsf{ABSTRACT}} \baselineskip=15pt
\medskip
It is well known that an asymptotically flat four-dimensional Kerr black hole has a smaller (specific) entropy than a Schwarzschild black hole of the same mass. We show here that the same is true if the temperature, rather than the mass, is held fixed; and we also show that an asymptotically AdS$_5$-Kerr black hole has a smaller specific entropy than an AdS$_5$-Schwarzschild black hole of the same temperature, except in a negligibly small class of special examples. The AdS$_5$-Kerr case is particularly interesting, because here the gauge-gravity duality applies; if we further accept that there is a useful analogy between the strongly coupled field theories dual to AdS black holes and the best-understood example of a strongly coupled fluid (the Quark-Gluon Plasma), then we can apply QGP theory to predict the behaviour of black hole entropy in this case. The prediction agrees with our study of AdS$_5$-Kerr entropy. The hope is that such results might lead ultimately to an identification of black hole microstates.

\newpage

\addtocounter{section}{1}
\section* {\large{\textsf{1. Rotation and Black Hole Entropy }}}
The celebrated Kerr black hole in the Cygnus X-1 system has many interesting properties. One of the less-discussed of these properties is that the specific entropy (that is, the entropy \cite{kn:wall,kn:thermong} per unit mass) of this black hole has only slightly more than \emph{half} of the value it would have if it were a Schwarzschild black hole of the same mass. This can be deduced from the formula for the specific entropy of a Kerr black hole (see below) and from the fact that the angular momentum of the Cygnus X-1 black hole is close to the maximum value permitted by Cosmic Censorship: the angular momentum per mass squared (in Planck units), $a_*,$ which is unity for an exactly extremal Kerr black hole, is claimed in \cite{kn:cygnus} to be at least\footnote{A more recent analysis \cite{kn:henric} suggests slightly smaller, but still very large values, in the $0.86$ to $0.92$ range. Of course, this black hole is by no means unique: for example, the black hole associated with the GW200129 gravitational waves is thought \cite{kn:hannam} to have a value of $a_*$ which may exceed 0.9.} $0.9696$ for Cygnus X-1 .

This is not a small effect. Astrophysical black holes notoriously have very large entropies \cite{kn:roger}, even when their large masses are taken into account. (This last point is one reason why, throughout our discussion below, we focus on the \emph{specific} entropy, and not the entropy itself.) This means that the rapid rotation of black holes like the one in Cygnus X-1 not only reduces the entropy, it reduces it by an amount which is very substantial even by astrophysical standards.

The rapid rotation in itself is not a mystery: there are well-developed detailed theories of the formation of such black holes (see for example \cite{kn:ying}), explaining why their angular momenta are so large, and the final total entropy is of course still (much) larger than the entropy of the initial system, in agreement with the Second Law.

What \emph{is} mysterious is that rotation should be \emph{in any way} related to the black hole entropy\footnote{We note in passing that this phenomenon plays an important role in applications of the Second Law of thermodynamics to black hole fragmentation \cite{kn:105} (because the fact that extremal black holes have the smallest possible entropy among black holes of a given mass allows one to put a lower bound on the entropy of the fragments), and it is fundamental in studies of the Penrose bound \cite{kn:khod}.}. It is true that the ``rotation'' of a black hole is actually a rather subtle matter; a black hole is not a physical object, so it does not ``rotate'' in the usual sense. Even leaving this to one side, a black hole does not rotate as if it were a rigid body ---$\;$ for example, when the two horizons are distinct they have different angular velocities \cite{kn:kerr}. Even taking these subtleties into account, however, it remains far from clear why this admittedly complex ``rotation'' should affect the specific entropy.

The reader may protest that this mystery is just one of many resulting from the fact that the microstates associated with black hole entropy are not well understood. We prefer to turn this around: if we could understand (even in special cases) why rotation reduces black hole (specific) entropy, then this might yield insights which could prove useful to the search for those microstates\footnote{For example, one can ask this question in the context of the recent remarkable results discussed in \cite{kn:chakra,kn:vijay}.}.

The reader might also protest: why make comparisons at fixed \emph{mass}? Would it not be more natural to compare the specific entropies of rotating and non-rotating black holes at the same \emph{temperature}? In this work, we will do this; and we find that it does not change the above discussion in its essentials: in the asymptotically flat case, a rotating black hole always has a lower specific entropy than a non-rotating black hole at the same temperature. (The asymptotically AdS case is more complex, but in nearly all cases the conclusion is the same.) The fixed-temperature case is actually the more interesting one for our purposes here, in particular for the (holographic) application to be discussed below (since temperature has a clear interpretation on both sides of the duality, and because the temperature is in fact approximately fixed when one makes comparisons using actual strongly coupled matter), so, after Section 3 below, we always take it as the basis of comparison.

One might try to explain the relation between rotation and entropy through a deep analysis of black hole entropy itself. However, even the most basic questions here have long been a matter of debate \cite{kn:tri}; even \emph{defining} an entropy for an intrinsically gravitational system is already notoriously difficult \cite{kn:gravent,kn:bin,kn:OYC}. Attempts to understand it (for example) in terms of densities of states lead to all manner of complications \cite{kn:DM} (see also \cite{kn:erik} for a recent discussion of the issues).

In view of all this, we prefer to focus on trying to understand our problem in a limited but still very instructive context: that of black holes with a \emph{holographic} interpretation.

\addtocounter{section}{1}
\section* {\large{\textsf{2. Rotation and Holography}}}
The gauge-gravity duality \cite{kn:casa,kn:nat,kn:bag,kn:noron,kn:edit} offers the possibility of approaching black hole entropy through a study of conformal field theories and related strongly coupled field theories. This approach has enjoyed several notable recent successes \cite{kn:CFT1,kn:CFT2,kn:CFT3} in \emph{computing} black hole entropies, and there is some hope \cite{kn:word} that it might in future be developed to the point where the precise duals of black hole microstates can be explicitly identified.

Some holographic formulations of the problem of identifying black hole microstates emphasise the possibility that the latter might perhaps be not very different from the microstates of other forms of strongly coupled matter. The concept of ``black hole molecules'' \cite{kn:mole} belongs to this category; see also more recent work along these lines summarised in \cite{kn:phonon,kn:thermong}.

In this spirit, it might be possible to understand the relation between black hole entropy and rapid rotation by studying another strongly coupled system in which ultra-high angular momentum to energy density ratios are to be found, namely the Quark-Gluon Plasma (QGP) produced in high energy collisions of nuclei \cite{kn:franc}. These plasmas exhibit attractor behaviour \cite{kn:solov} and constitute a well-defined thermodynamic system. When the collision is peripheral, that is, not ``head-on'', one can expect very large specific angular momenta to be generated, and this has been confirmed experimentally in the STAR observations \cite{kn:STARcoll} at the RHIC facility (see \cite{kn:sass} for up-to-date references). In this system, one can compare the ``vortical QGP'' with non-vortical plasmas produced in \emph{central} collisions in the same beam, that is, at the same impact energy (and, approximately, the same temperature).

These systems are relevant here because, as is well known, the real, strongly coupled QGP resembles, in some ways, the field theories appearing in gauge-gravity duality. (See \cite{kn:93,kn:yidian,kn:anast,kn:config} and their references for the application of holography to rotating strongly coupled matter). If we accept this resemblance as a heuristic guide, then the various physical parameters describing the QGP are mapped to analogous parameters of an asymptotically AdS$_5$ black hole: temperature to Hawking temperature, the plasma angular momentum/energy density ratio to the specific angular momentum\footnote{It is possible for an AdS$_5$-Kerr black hole to rotate about two different axes simultaneously, with two different specific angular momenta. Clearly this is of no use to us here, so we always take it that our black holes rotate about a single axis, with one specific angular momentum.} of an AdS$_5$-Kerr black hole, the QGP entropy/energy density ratio to the specific entropy of that black hole, and so on.

Now a comparison of the entropy densities of vortical and non-vortical plasmas produced in high-energy collisions can be made rather explicitly, as follows. The vorticities reported \cite{kn:STARcoll} for peripheral collisions are thought to be \cite{kn:ivansold1,kn:ivansold2} averages over small vortices produced in shearing layers. These small vortices tend to align with each other, due to spin-spin coupling of the angular momentum vectors of the underlying particles. This gives rise to two observable phenomena.

First, in the specific case of the decay of $\Lambda$/$\overline{\Lambda}$ hyperons produced in heavy ion collisions, protons are emitted along the direction of the spin of the parent hyperon, and this polarization effect is the one reported in the relevant STAR experiments.

The spin-spin coupling also gives rise to a magnetic field \cite{kn:chern2,kn:nica}, in close analogy to the classical \emph{Barnett effect} \cite{kn:barn}. These (very large) magnetic fields, arising in peripheral but not central collisions, have been studied extensively: see for example \cite{kn:vlad,kn:kaz,kn:chern1,kn:hao}. They can give rise to observable effects, and in many circumstances they can act as a proxy for the average vorticity, with which they are correlated.

The key point is this: the Barnett-effect alignments constrict the relevant phase space, reducing the entropy density/energy density ratio (or the entropy per particle) at a given temperature. Thus we expect the plasmas with high angular momentum densities to have smaller specific entropies than their counterparts produced in central collisions at the same impact energy, producing plasmas of approximately the same temperature.

In fact, phenomenological studies \cite{kn:hof} strongly suggest that a decline in entropy density with increasing magnetic fields (at fixed temperature) is indeed a quite general property of QCD thermodynamics, though this is only confirmed at low temperatures; however there are indications of similar phenomena occurring in the QGP (see for example \cite{kn:bali}). It is reasonable to expect that the same is true when the magnetic field is generated by the Barnett effect. The influence of rotation on the thermodynamics of the QGP remains a topic of great current interest \cite{kn:chernodub}, and perhaps this prediction can be confirmed using such methods.

As with any application of holography, one has to be aware of the limitations of as well as the insights gained from this method \cite{kn:mateos}. In the case of the usual AdS$_5$-Kerr black hole (with a topologically spherical event horizon), the dual system is not a simple vortex in a strongly coupled fluid: it is an entire rotating three-sphere. However, the Barnett effect involves the \emph{entropy per particle}, as we mentioned earlier. We can reasonably hope to capture this behaviour holographically by focusing exclusively on an intensive quantity, the \emph{specific} entropy, which by construction is insensitive to the sizes of the two systems. (The fact that the boundary space is curved should not be a problem, since \cite{kn:93} this curvature is quite small.) This is another reason for us to focus on this particular quantity: we do not necessarily expect to be able to model the entropy itself in this manner.

To the extent that the real QGP mirrors the properties of field theories that are dual to AdS$_5$ bulk black holes ---$\,$ or, to put it differently, if strongly coupled matter is always subject to the analogue of the Barnett effect ---$\,$ holographic duality then predicts that rotating AdS$_5$ black holes should have lower specific entropies than their non-rotating counterparts at the same temperature.

As was mentioned briefly in the preceding section, in this work we will confirm this prediction by showing that rotation does (nearly) always reduce the specific entropy of asymptotically AdS$_5$ black holes, when the temperature is held fixed as the basis of comparison. To be precise, we find that, for extremely small values of the specific angular momentum, the specific entropy (in the case of so-called ``large'' AdS$_5$ black holes) actually rises at first, but \emph{only} when the temperature is carefully adjusted to a value close to the minimal possible value for an asymptotically AdS$_5$-Schwarzschild black hole\footnote{Lower temperatures are possible by beginning with a black hole that is already rotating; but then no comparison can be made with a corresponding non-rotating black hole. Nevertheless we will briefly consider this possibility later, finding that it does not contradict the statements being made here.}. Furthermore, even when it occurs at all, this effect is very quickly reversed as the specific angular momentum increases. \emph{Generically, then, the tendency is for the specific entropy of an AdS$_5$ black hole to decrease as the specific angular momentum increases, at fixed temperature.}

To summarize this whole discussion: in the concrete example of the QGP produced in heavy ion collisions, one has a rather clear statistical-mechanical picture, in terms of physics similar to that underlying the Barnett effect (and its generalization \cite{kn:chern2,kn:nica} to the QGP), of the decrease of the ratio of the entropy and energy densities as we move from central to peripheral collisions in the same beam (with temperature therefore approximately fixed). If this is a universal property of strongly coupled matter, holography then predicts the decrease of the specific entropy of AdS$_5$-Kerr black holes as their specific angular momenta are increased to large values at fixed temperature. The main technical objective of this work is to show that, with minor exceptions, this is in fact the case. In this sense, we have an explanation, admittedly only in the AdS case, of the fact that rotation affects (in fact, reduces) black hole specific entropy.

While our primary concern here is with rotating black holes, electromagnetically charged black holes are also of interest and often throw light on the (much) more intricate rotating case. These Reissner-Nordstr\"{o}m and AdS$_5$-Reissner-Nordstr\"{o}m black holes were long regarded as unphysical, but they have recently undergone a revival of interest: see for example \cite{kn:hoop,kn:hold}. In particular, the \emph{magnetically} charged case has attracted considerable attention \cite{kn:yang,kn:ullah}. Again, electromagnetic charges reduce the specific entropy (at either fixed mass or fixed temperature). Indeed, these black holes are expected \cite{kn:juan1} generically to be close to extremality, and an extremal Reissner-Nordstr\"{o}m black hole has an even smaller specific entropy than an extremal Kerr black hole of the same mass: in the asymptotically flat case it has only \emph{one quarter} of the specific entropy of the corresponding Schwarzschild black hole.

We will consider three cases: asymptotically flat, four-dimensional black holes at fixed mass ---$\,$ this because the details are so simple that the issues can be seen clearly and explicitly ---$\,$ and then asymptotically flat, four-dimensional black holes at fixed temperature, and finally the case of real interest to us, asymptotically AdS, five-dimensional black holes at fixed temperature, which are technically considerably more challenging.

Before we begin, note that in all cases we will be computing entropies from the areas of event horizons, which of course means that we assume that those horizons exist: that is, we assume that Cosmic Censorship holds in all cases. In the strict sense, this can be disputed, but it now seems likely \cite{kn:roberto} that if naked singularities can occur, they are severely constricted in both space and time; that is, they arise in certain very limited regions of highly dynamical spacetimes, for example when black holes collide or bifurcate. This is probably true even in the asymptotically AdS case \cite{kn:newemp}. Since we are only interested in non-dynamical spacetimes here, we anticipate no difficulties on this score, and so we assume Censorship henceforth.

\addtocounter{section}{1}
\section* {\large{\textsf{3. Four-Dimensional, Asymptotically Flat, Fixed Mass}}}
The First Law of black hole thermodynamics \cite{kn:wall} takes the familiar form\footnote{Because we will be discussing both four- and five-dimensional spacetimes, we avoid Planck units, and use units in which mass and temperature are inverse lengths, the four-dimensional Coulomb constant is dimensionless and set equal to unity (but the five-dimensional Coulomb constant is \emph{not} ---$\,$ it has units of length) and entropy, angular momentum, and charge are dimensionless (so specific entropies and angular momenta have units of length). We denote the specific angular momentum in these units by $\mathfrak{j};$ we reserve the notation $a$ for the angular momentum parameter which occurs in the various Kerr metrics.}
\begin{equation}\label{A}
d\mathcal{M}\;=\;TdS\,+\,\Omega d\mathcal{J} \,+\, \Phi d\mathcal{Q},
\end{equation}
where $T$ is the Hawking temperature, $S$ is the black hole entropy, $\Omega$ is the angular velocity of a light ray skimming the horizon, $\Phi$ is the potential at the horizon, $\mathcal{M}$ is the black hole mass, $\mathcal{J}$ is the black hole angular momentum, and $\mathcal{Q}$ is the charge. It is clear that, if we increase the charge or the angular momentum while fixing the mass, then $dS < 0;$ the entropy (and likewise the specific entropy) has to be smaller.

This statement is readily made explicit. For example, an asymptotically flat four-dimensional Reissner-Nordstr\"{o}m black hole with mass $\mathcal{M},$ charge $\mathcal{Q}$, and metric
\begin{flalign}\label{B}
g\left(\textsf{AFRN}_4\right)\;=\;&-\,\left(1\,-\,{2\mathcal{M}\ell_4^2\over r}\,+\,{\mathcal{Q}^2\ell_4^2\over 4\pi r^2}\right)\m{d}t^2\,+{\m{d}r^2\over 1\,-\,{2\mathcal{M}\ell_4^2\over r}\,+\,{\mathcal{Q}^2\ell_4^2\over 4\pi r^2}}\\ \notag
\,\,\,\,&\,+\,r^2\left(\m{d}\theta^2 \,+\, \sin^2\theta\,\m{d}\phi^2\right),
\end{flalign}
has a specific entropy given by
\begin{equation}\label{C}
\mathfrak{s}_{\textsf{AFRN}_4} \;=\;{\pi r_{\textsf{H}}^2\over \ell_4^2\mathcal{M}} \;=\; \pi \mathcal{M}\ell_4^2\,\left[1\;+\;\left(1\;-\;{\mathcal{Q}^2\over4\pi \ell_4^2\mathcal{M}^2}\right)\;+\;2\sqrt{1\;-\;{\mathcal{Q}^2\over4\pi \ell_4^2\mathcal{M}^2}}\,\right].
\end{equation}
Here $r_{\textsf{H}}$ is the radial coordinate at the event horizon, and $\ell_4$ is the four-dimensional Planck length. (The extremal case is $\mathcal{Q}^2 = 4 \pi\ell_4^2\mathcal{M}^2,$ so indeed, as claimed above, the (specific) entropy in that case is one quarter of the corresponding Schwarzschild specific entropy, $4\pi \mathcal{M}\ell_4^2.$)

It is obvious in this case that the specific entropy is indeed a monotonically decreasing function of the charge, assuming the mass to be fixed. The point is simply that adding charge causes the event horizon to contract (for fixed mass) and of course this reduces the specific entropy. This does not \emph{explain} the reduction in statistical-mechanical terms, but at least one has a clear intuitive picture of it.

Let us now consider the asymptotically flat four-dimensional Kerr metric,
\begin{flalign}\label{D}
g\left(\textsf{AFK}_4\right) = &- {\Delta_r \over \rho^2}\left[\,\m{d}t \; - \; \mathfrak{j}\,\sin^2\theta \,\m{d}\phi\right]^2\;+\;{\rho^2 \over \Delta_r}\m{d}r^2\;+\;\rho^2\,\m{d}\theta^2 \\ \notag \,\,\,\,&+\;{\sin^2\theta \, \over \rho^2}\left[\mathfrak{j}\,\m{d}t \; - \;\left(r^2\,+\,\mathfrak{j}^2\right)\,\m{d}\phi\right]^2,
\end{flalign}
where
\begin{eqnarray}\label{E}
\rho^2& = & r^2\,+\,\mathfrak{j}^2\cos^2\theta, \nonumber\\
\Delta_r & = & r^2+\mathfrak{j}^2 - 2\mathcal{M}\ell_4^2r,
\end{eqnarray}
and where $\mathcal{M}$ is the mass, and $\mathfrak{j}$ is the specific angular momentum, of the black hole. (The reason for the slightly unconventional notation here will be explained later.)

The effect of rotation on the entropy is less clear here than in the Reissner-Nordstr\"{o}m case, because rotation distorts the event horizon, causing it to become an oblate spheroid; the area is now $4\pi \left(r_{\textsf{H}}^2 + \mathfrak{j}^2\right)$, where $r_{\textsf{H}}$ denotes the location of the event horizon. If $r_{\textsf{H}}$ remained constant, then clearly rotation would \emph{increase} the specific entropy, because of the presence of the $\mathfrak{j}^2$ term in the formula for the area.

The resolution is that, once again, rotation causes the radius to become smaller for fixed mass, and this effect is \emph{larger than} and overcomes the effect of the distortion of the event horizon; but at a qualitative level it is not clear that this is so. It can only be seen from the explicit formula for the specific entropy in this case:
\begin{equation}\label{F}
\mathfrak{s}_{\textsf{AFK}_4} \;=\;{\pi \left(r_{\textsf{H}}^2 + \mathfrak{j}^2\right)\over \ell_4^2\mathcal{M}} \;=\; 2\pi \mathcal{M}\ell_4^2\left[1 \;+\;\sqrt{1\;-\;{\mathfrak{j}^2\over \ell_4^4\mathcal{M}^2}}\,\right].
\end{equation}
Again, the specific entropy decreases with $\mathfrak{j}$ for fixed mass, as the First Law requires. But we see that the reduction is smaller than in the Reissner-Nordstr\"{o}m situation: in the extremal case, $\mathfrak{j}^2 = \ell_4^4\mathcal{M}^2,$ the specific entropy is one half of the value $4\pi \mathcal{M}\ell_4^2$ for a Schwarzschild black hole of the same mass, not one quarter.

In summary: in the Kerr case, rotation has \emph{two}, opposite, effects on the specific entropy, because it affects both the size \emph{and} the shape of the black hole. Unless the First Law is invoked, it is not immediately clear which effect will prevail. Clearly, the rotating case is considerably more subtle than its charged counterpart.

Next we turn to the case in which the temperatures are fixed, rather than the masses.

\addtocounter{section}{1}
\section* {\large{\textsf{4. Four-Dimensional, Asymptotically Flat, Fixed Temperature}}}
In this section, we repeat the computations above, but now keeping the Hawking temperature fixed as we vary the charge or angular momentum. This means that we must avoid expressions involving the mass explicitly, since the mass will vary in some way that is hard to control. This in turn means that we have no guarantee from the First Law that the specific entropy will necessarily decrease with increasing charge or angular momentum.

For the remainder of this work, we never allow the Hawking temperature to be zero; that is, we exclude extremal black holes. The reasons for this exclusion will be explained later, when we consider our primary interest, AdS$_5$-Kerr black holes. Henceforth, therefore, we permit ourselves to divide by the Hawking temperature where necessary.

We begin with the charged case.

The Hawking temperature of a four-dimensional, asymptotically flat Reissner-Nordstr\"{o}m black hole can be put into the form
\begin{equation}\label{G}
T_{\textsf{AFRN}_4}\;=\;{1\over 4\pi r_{\textsf{H}}}\;-\;{\mathcal{Q}^2\ell_4^2\over 16\pi^2r_{\textsf{H}}^3},
\end{equation}
where, as explained, we have eliminated the mass, using the definition of $r_{\textsf{H}}$. Similarly, the specific entropy can be expressed in a way such that the mass does not appear:
\begin{equation}\label{H}
\mathfrak{s}_{\textsf{AFRN}_4}\; = \; {2\pi r_{\textsf{H}} \over 1\;+\; {\mathcal{Q}^2\ell_4^2 \over 4\pi r_{\textsf{H}}^2}}
\end{equation}

It is easily shown that, if we fix $T_{\textsf{AFRN}_4}$ and then use (\ref{G}) to regard $r_{\textsf{H}}$ as a function of $\mathcal{Q},$ then it is a decreasing function. Consequently, (\ref{H}) implies that $\mathfrak{s}_{\textsf{AFRN}_4}$ is still a decreasing function of the charge, just as it was in the preceding section.

More explicitly: solving (\ref{G}) for $r_{\textsf{H}}$, and substituting the result into (\ref{H}), one obtains an expression for $\mathfrak{s}_{\textsf{AFRN}_4}$ in terms of $T_{\textsf{AFRN}_4}$ (which we abbreviate to $T$) and $\mathcal{Q}$:

\bigskip

$\mathfrak{s}_{\textsf{AFRN}_4}\; =$
\begin{equation}\label{I}
 { {1\over 6 T}\left(\,\sqrt [3]{-54\,\pi \,{\mathcal{Q}}^{2}{T}^{2}{\ell_4}^{2}+6\,\sqrt {3}\mathcal{Q}\ell_4\sqrt {
{\frac {27\,\pi \,{\mathcal{Q}}^{2}{T}^{2}{\ell_4}^{2}-1}{\pi }}}\pi \,T+1}\;+\;{\frac {1}{\sqrt [3]{-54\,\pi \,{\mathcal{Q}}
^{2}{T}^{2}{\ell_4}^{2}+6\,\sqrt {3}\mathcal{Q}\ell_4\sqrt {{\frac {27\,\pi \,{\mathcal{Q}}^{2}{T}^{
2}{\ell_4}^{2}-1}{\pi }}}\pi \,T+1}}}+1\right)\over 1 \;+\; {144\pi^2 T^2\mathcal{Q}^2\ell_4^2\over \left(\,\sqrt [3]{-54\,\pi \,{\mathcal{Q}}^{2}{T}^{2}{\ell_4}^{2}+6\,\sqrt {3}\mathcal{Q}\ell_4\sqrt {
{\frac {27\,\pi \,{\mathcal{Q}}^{2}{T}^{2}{\ell_4}^{2}-1}{\pi }}}\pi \,T+1}\;+\;{\frac {1}{\sqrt [3]{-54\,\pi \,{\mathcal{Q}}
^{2}{T}^{2}{\ell_4}^{2}+6\,\sqrt {3}\mathcal{Q}\ell_4\sqrt {{\frac {27\,\pi \,{\mathcal{Q}}^{2}{T}^{
2}{\ell_4}^{2}-1}{\pi }}}\pi \,T+1}}}+1\right)^2}}
\end{equation}

One sees that the level of complexity here has risen very substantially (compare with (\ref{C})). It is by no means clear that the function on the right side of (\ref{I}) should have any particular monotonicity property. But, as we know, it does: for all fixed $T_{\textsf{AFRN}_4}$, $\mathfrak{s}_{\textsf{AFRN}_4}$ is monotonically decreasing on its domain (which is determined by the fact that there is a (mass-independent) upper bound\footnote{This upper bound is given by $$|\mathcal{Q}| \leq {1\over 3\sqrt{3\pi}\,\ell_4T_{\textsf{AFRN}_4}}.$$} on the charge when the temperature is fixed). For example, if we (temporarily) use units in which $\ell_4 = 1$ and set $T_{\textsf{AFRN}_4} = 0.25,$ then the graph of $\mathfrak{s}_{\textsf{AFRN}_4}$ is shown as Figure 1.

\begin{figure}[!h]
\centering
\includegraphics[width=0.6\textwidth]{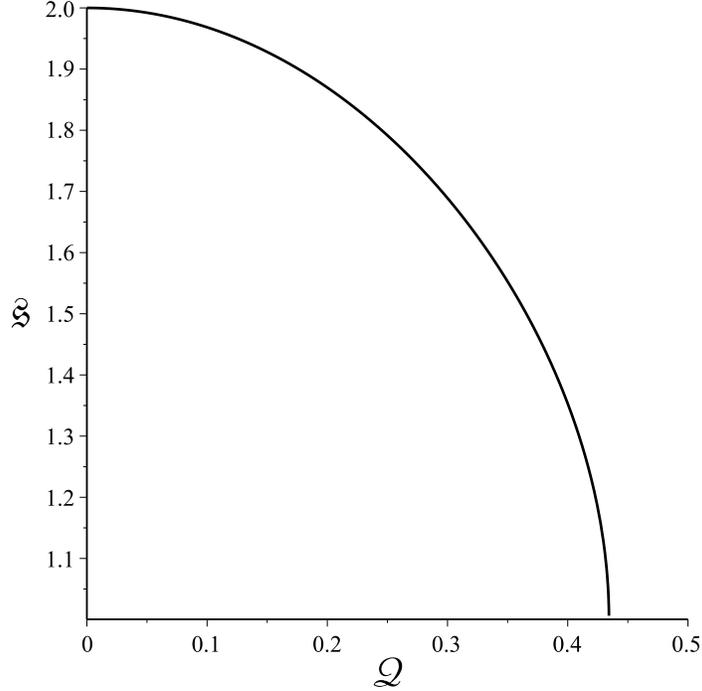}
\caption{Specific entropy of a four-dimensional asymptotically flat Reissner-Nordstr\"{o}m black hole with fixed temperature 0.25 in Planck units, as a function of the charge.}
\end{figure}

A straightforward calculation shows that $r_{\textsf{H}}$ is bounded below by $1/(6\pi T_{\textsf{AFRN}_4}),$ and that the specific entropy is bounded below by $1/(4T_{\textsf{AFRN}_4})$, which is exactly \emph{one half} of the value when $\mathcal{Q} = 0$. This contrasts sharply with the minimal value (one quarter) in the preceding section. That is, when the temperature is fixed rather than the mass, the specific entropy is still reduced by increasing the charge, but \emph{much less effectively}.

Turning now to the Kerr case: the temperature is \cite{kn:hawk,kn:cognola,kn:gibperry}
\begin{equation}\label{J}
T_{\textsf{AFK}_4}\;=\;{r_{\textsf{H}}\left(1 - {\mathfrak{j}^2\over r_{\textsf{H}}^2}\right)\over 4\pi \left(\mathfrak{j}^2 + r_{\textsf{H}}^2\right)}.
\end{equation}
One can show that, if we use this to think of $r_{\textsf{H}}$ as a function of $\mathfrak{j}$ for fixed $T_{\textsf{AFK}_4}$, then it is a decreasing function.

As before, we wish to express the specific entropy in a manner that does not involve $\mathcal{M}$ explicitly. To that end, we set $\Delta_r = 0$ in (\ref{E}) and use that equation to express $\mathcal{M}$ in terms of $r_{\textsf{H}}$ and $\mathfrak{j}$. The result is surprisingly simple:
\begin{equation}\label{K}
\mathfrak{s}_{\textsf{AFK}_4} \;=\;{\pi \left(r_{\textsf{H}}^2 + \mathfrak{j}^2\right)\over \ell_4^2\mathcal{M}} \;=\;2\pi r_{\textsf{H}}.
\end{equation}
In this case there is no ``competition'', and so it is immediate that the effect of increasing the specific angular momentum while fixing the temperature is to cause the specific entropy to decrease.

Explicitly, and again abbreviating $T_{\textsf{AFK}_4}$ to $T$, we have
\begin{eqnarray}\label{L}
\mathfrak{s}_{\textsf{AFK}_4}\;& =\; & {1\over 6T}\, \left( \sqrt [3]{-288\,{\pi }^{2}{T}^{2}{\mathfrak{j}}^{2}+12\,\sqrt {3}
\sqrt {256\,{\pi }^{4}{T}^{4}{\mathfrak{j}}^{4}+176\,{\pi }^{2}{T}^{2}{\mathfrak{j}}^{2}\;-1}\;
\pi \,T\,\mathfrak{j}\,+1}\right. \\ \notag
& &\left.-{\frac {48\,{\pi }^{2}{T}^{2}{\mathfrak{j}}^{2}-1}{\sqrt [3]{-288\,{
\pi }^{2}{T}^{2}{\mathfrak{j}}^{2}+12\,\sqrt {3}\sqrt {256\,{\pi }^{4}{T}^{4}{\mathfrak{j}}
^{4}+176\,{\pi }^{2}{T}^{2}{\mathfrak{j}}^{2}-1}\;\pi \,T\,\mathfrak{j}\,+1}}}+1 \right).
\end{eqnarray}
Similarly to the Reissner-Nordstr\"{o}m case, here there is a (mass-independent) upper bound\footnote{This upper bound is given by $$\mathfrak{j} \leq {\sqrt{2}\left(\sqrt{5} - 1\right)^{5/2} \over 32\pi T_{\textsf{AFK}_4}}$$} on $\mathfrak{j}$, for fixed $T_{\textsf{AFK}_4}$. This corresponds to a lower bound on the specific entropy itself, given by
\begin{equation}\label{M}
\mathfrak{s}_{\textsf{AFK}_4} \;\geq\;{\sqrt{5} - 1\over 4T_{\textsf{AFK}_4}}\;\approx {0.309 \over T_{\textsf{AFK}_4}}.
\end{equation}
This means that the specific entropy can in this case never be smaller than approximately $61.8\%$ of the specific entropy of a non-rotating black hole of the same temperature (which is $1/(2T_{\textsf{AFK}_4})$). As in the Reissner-Nordstr\"{o}m case, angular momentum is less effective in reducing the specific entropy when the temperature is fixed than when the mass is fixed.

For an example illustrating these points, if we use four-dimensional Planck units and set the temperature at $0.1,$ then the graph of $\mathfrak{s}_{\textsf{AFK}_4}$ is shown as Figure 2.

\begin{figure}[!h]
\centering
\includegraphics[width=0.6\textwidth]{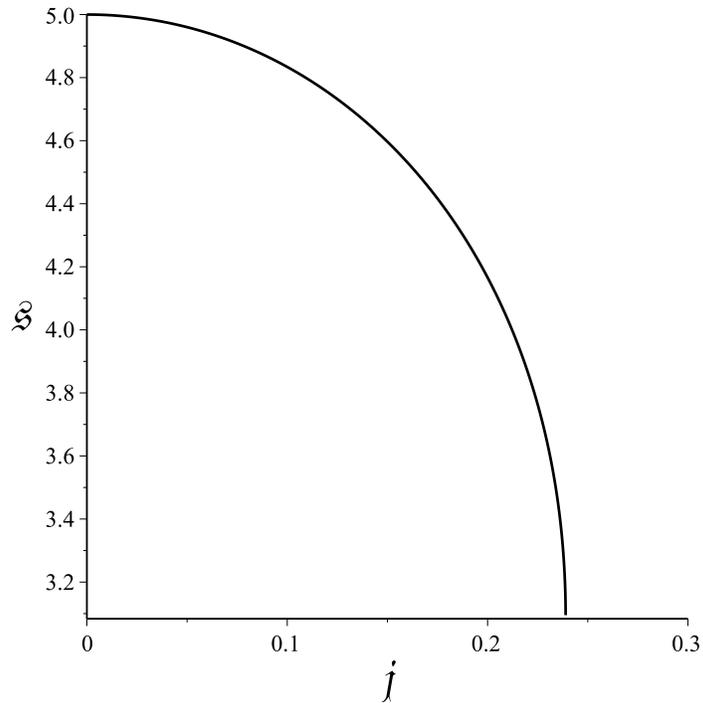}
\caption{Specific entropy of a four-dimensional asymptotically flat Kerr black hole with fixed temperature 0.1 in Planck units, as a function of the specific angular momentum.}
\end{figure}

With this preparation, we turn to the case of five-dimensional asymptotically AdS$_5$ black holes, with a fixed temperature.

\addtocounter{section}{1}
\section* {\large{\textsf{5. Five-Dimensional, Asymptotically AdS$_5$, Fixed Temperature}}}
With a view to applying the gauge-gravity duality, let us now turn to the asymptotically AdS, five-dimensional case. (For simplicity, we will assume that the topology of the event horizon, in both cases to be considered, is that of the three-sphere. See for example \cite{kn:108} for a recent discussion of other choices. For other aspects of the thermodynamics of AdS black holes, see for example \cite{kn:man}.)

For reasons to be explained, it turns out that the AdS$_5$-Kerr case is very intricate. We therefore begin with the asymptotically AdS$_5$-Reissner-Nordstr\"{o}m black hole, which is much simpler and yet surprisingly similar to the rotating case in practice.

\addtocounter{subsection}{1}
\section* {\large{\textsf{5.1. AdS$_5$-Reissner-Nordstr\"{o}m, Fixed Temperature}}}
The metric in this case is
\begin{flalign}\label{N}
g(\textsf{AdSRN}_5)\;=\;&-\,\left(1\,+\,{r^2\over L^2}\,-\,{8\mathcal{M}\ell_5^3\over 3\pi r^2}\,+\,{k_5\mathcal{Q}^2\ell_5^3\over 3\pi^3 r^4}\right)\m{d}t^2\,+{\m{d}r^2\over 1\,+\,{r^2\over L^2}\,-\,{8\mathcal{M}\ell_5^3\over 3\pi r^2}\,+\,{k_5\mathcal{Q}^2\ell_5^3\over 3\pi^3 r^4}}\\ \notag \,\,\,\,&\,+\,r^2\left(\m{d}\theta^2 \,+\, \sin^2\theta\,\m{d}\phi^2\,+\,\cos^2\theta\,\m{d}\psi^2\right).
\end{flalign}
Here $\mathcal{M}$ and $\mathcal{Q}$ are the physical mass and charge, $L$ is the asymptotic AdS$_5$ curvature length scale, $\ell_5$ is the AdS$_5$ Planck length, $k_5$ is the five-dimensional Coulomb constant (with units of length, as mentioned earlier), and the coordinates on the three-sphere are Hopf coordinates\footnote{These differ from spherical polar coordinates: $\phi$ and $\psi$ run from $0$ to $2\pi,$ but $\theta$ runs from $0$ to $\pi/2$.}.

The Hawking temperature can be expressed as
\begin{equation}\label{O}
T_{\textsf{AdSRN}_5}\;=\;{1\over 4\pi}\,\left[{4r_{\textsf{H}}\over L^2}\;+\;{2\over r_{\textsf{H}}}\;-\;{2k_5\mathcal{Q}^2\ell_5^3\over 3\pi^3r_{\textsf{H}}^5}\right],
\end{equation}
where $r_{\textsf{H}}$ locates the event horizon as before. The specific entropy is
\begin{equation}\label{P}
\mathfrak{s}_{\textsf{AdSRN}_5}\; = \; {\pi^2r_{\textsf{H}}^3\over 2\ell_5^3\mathcal{M}}\;=\;{4\pi r_{\textsf{H}}\over 3\left(1 + {r_{\textsf{H}}^2\over L^2} + {k_5\mathcal{Q}^2\ell_5^3\over 3 \pi^3 r_{\textsf{H}}^4}\right)}.
\end{equation}
One can usefully write this, by using equation (\ref{O}), as
\begin{equation}\label{Q}
\mathfrak{s}_{\textsf{AdSRN}_5}\;=\;{4\pi r_{\textsf{H}}\over 3\left(2 + {3r_{\textsf{H}}^2\over L^2} - 2\pi T_{\textsf{AdSRN}_5}r_{\textsf{H}} \right)}.
\end{equation}

Now recall that AdS$_5$-Schwarzschild black holes cannot have arbitrarily small temperatures: the minimal temperature is $\sqrt{2}/\left(\pi L\right)$. This is thought \cite{kn:edwit} to reflect holographically the fact that strongly coupled matter at zero baryonic chemical potential cannot exist below a certain temperature: it confines. For any temperature above $\sqrt{2}/\left(\pi L\right)$, there are two such black holes, one with a larger event horizon than the other; one speaks of ``small'' and ``large'' black holes. The ``large'' black holes are the ones that actually describe strongly coupled matter holographically, since (at these temperatures) they have lower action and more conventional thermodynamic behaviour (notably, positive specific heat) than asymptotically flat black holes. In addition, they are able to attain equilibrium \cite{kn:ruong}. This reflects the actual thermodynamics of the equivalent boundary matter, which we hope to use to study the behaviour of the dual black hole.

Henceforth, then, we mainly confine attention to AdS$_5$ black holes which, when they have no charge or angular momentum, are ``large''. We then study the effects of adding charge or angular momentum to these black holes, while fixing the temperature as usual. This means that (nearly) all of our black holes have temperature greater than or equal to $\sqrt{2}/\left(\pi L\right)$; in particular, it means that we do not consider extremal black holes, as has been our understanding in all of our discussions thus far. (AdS$_5$-Kerr black holes with temperatures less than $\sqrt{2}/\left(\pi L\right)$ will however be discussed briefly below, when we discuss the possible superradiant instability of such black holes.)

We begin by observing that we can use equation (\ref{O}) to regard $r_{\textsf{H}}$ as a function of $\mathcal{Q},$ while fixing the temperature. Ideally we would like to solve (\ref{O}) for $r_{\textsf{H}}$ and then use the result to express $\mathfrak{s}_{\textsf{AdSRN}_5}$ as a function of $\mathcal{Q}$. Unfortunately that is not feasible (since (\ref{O}) is a sextic in $r_{\textsf{H}}$), so we proceed more indirectly, by studying the nature of the relationship between $r_{\textsf{H}}$ and $\mathcal{Q}.$

If we fix the temperature at some value above $\sqrt{2}/\left(\pi L\right)$, and graph $r_{\textsf{H}}$ as a function of $\mathcal{Q},$ then we find that the graph consists of two disconnected parts, corresponding on the vertical axis to ``large'' and ``small'' AdS$_5$-Schwarzschild black holes. The branch of the graph corresponding to the ``large'' black hole of some given temperature is of course the upper one: see for example Figure 3.

\begin{figure}[!h]
\centering
\includegraphics[width=0.6\textwidth]{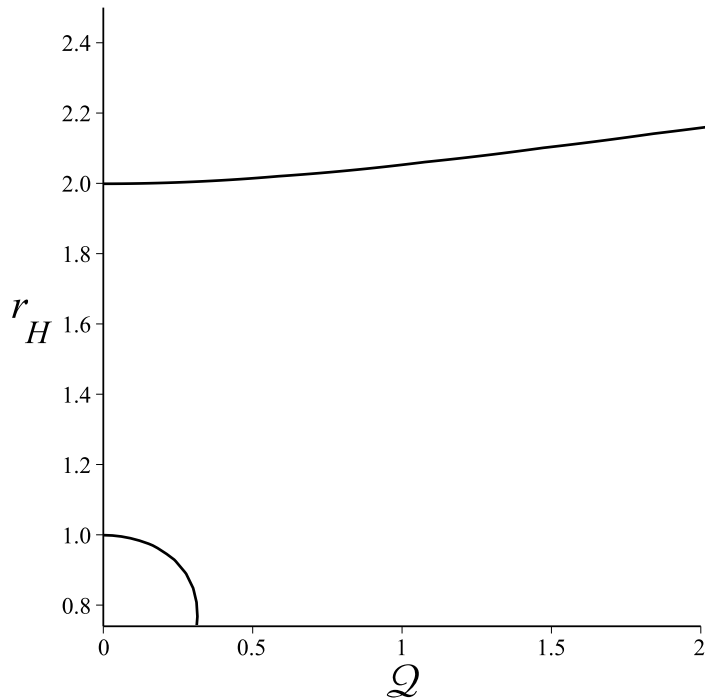}
\caption{Radial coordinate at horizon for AdS$_5$-Reissner-Nordstr\"{o}m black holes, as a function of the charge. (Parameters, including the temperature, fixed in units with $L = 1,$ at values chosen for illustrative clarity.) The corresponding ``large'' AdS$_5$-Schwarzschild black hole has $r_{\textsf{H}} = 2$ in these units. }
\end{figure}

We see now that adding electric charge to a ``large'' AdS$_5$-Schwarzschild black hole, while keeping its temperature fixed, actually \emph{increases}\footnote{One also sees from Figure 3 that ``small'' black holes behave, as expected, like their asymptotically flat counterparts: that is, the event horizon contracts with increasing charge. As in the asymptotically flat case, there is a mass-independent upper bound on the charge of ``small'' black holes of fixed temperature (but not in the ``large'' case).} $r_{\textsf{H}}$. Note that there is no upper bound on $r_{\textsf{H}}$; it can attain any value above the minimum if the charge is sufficiently large. (If $r_{\textsf{H}}$ were bounded above as $\mathcal{Q}$ approached infinity, there would be a contradiction with equation (\ref{O}).)

We now turn to the study of the specific entropy (confining attention to the ``large'' case). To see what happens to it when the charge is increased, note first that, as we have seen, $r_{\textsf{H}}$ is an increasing, unbounded function of the charge; and so, to study the effect of increasing charge on the specific entropy, we can study instead the effect of increasing $r_{\textsf{H}}$.

From equation (\ref{Q}), we see that when the specific entropy is regarded as a function of $r_{\textsf{H}}$, it actually increases, reaching a maximum value at a certain value of $r_{\textsf{H}}$: see for example Figure 4. A straightforward calculation shows that this value is \emph{always} (that is, for all temperatures) $\sqrt{2/3}\,L \approx 0.816 L.$ However, this in itself does not mean that the specific entropy can increase with increasing charge: we need to determine whether the physical domain for $r_{\textsf{H}}$ begins to the left or to the right of that maximum point.

As we have seen, in the ``large'' case the smallest possible value of $r_{\textsf{H}}$ at any fixed temperature $T_{\textsf{AdSRN}_5}$ is given by its value when the charge is zero, so the left end of the physical domain for $r_{\textsf{H}}$ is determined by its value for a ``large'' AdS$_5$-Schwarzschild black hole of that temperature. That value, for any temperature $T$ above the minimum, is given by
\begin{equation}\label{QMINUS}
r_{\textsf{H}}^{\textsf{AdS}_5\textsf{Large Sch}}\;=\;{L^2\over 2}\,\left(\pi T + \sqrt{\pi^2 T^2 - (2/L^2)}\right).
\end{equation}

Consider first the case when this temperature takes the lowest possible value for an AdS$_5$-Schwarzschild black hole; as was mentioned earlier (and as can be seen by inspecting (\ref{QMINUS})), that value is $\sqrt{2}/\left(\pi L\right)$. In that case, $r_{\textsf{H}} = L/\sqrt{2} \approx 0.707 L,$ which is indeed smaller than the value at the maximum. It follows that, at this temperature, the specific entropy actually \emph{increases} as the charge increases from zero. However, this is only true for very small charges: as soon as $r_{\textsf{H}}$ increases from $\approx 0.707 L$ and reaches $\approx 0.816L,$ the specific entropy begins to decline: see Figure 4.

\begin{figure}[!h]
\centering
\includegraphics[width=0.6\textwidth]{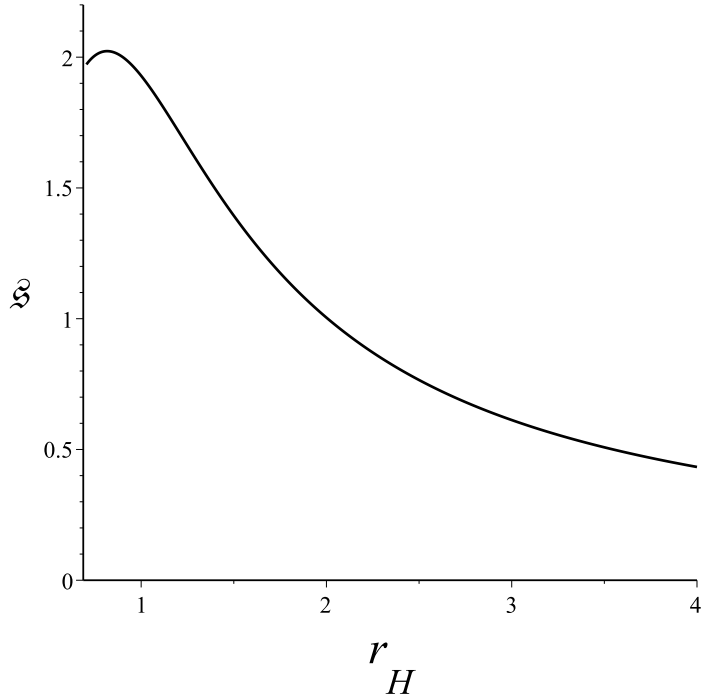}
\caption{Specific entropy of a ``large'' AdS$_5$-Reissner-Nordstr\"{o}m black hole with temperature equal to the minimal possible temperature of an AdS$_5$-Schwarzschild black hole, using units in which $L = 1$. Here $r_{\textsf{H}}$ is a proxy for the charge, with which it varies as a monotonically increasing function when the temperature is fixed. The physical domain is to the right of $r_{\textsf{H}} \approx 0.707,$ which is slightly smaller than at the maximum, $r_{\textsf{H}} \approx 0.816$. }
\end{figure}

Furthermore, this only happens when the temperature is just above the minimal possible temperature of an AdS$_5$-Schwarzschild black hole. To see this, we simply set $r_{\textsf{H}} = \sqrt{2/3}\,L,$ since that is the location of the maximum. One finds easily that $r_{\textsf{H}}$ in the AdS$_5$-Schwarzschild case attains this value when the temperature is $7\sqrt{3}/12 \,\approx 1.010$ times the minimal possible temperature. Beyond this temperature, the specific entropy always decreases with $r_{\textsf{H}}$ on its physical domain: see Figure 5.

\begin{figure}[!h]
\centering
\includegraphics[width=0.6\textwidth]{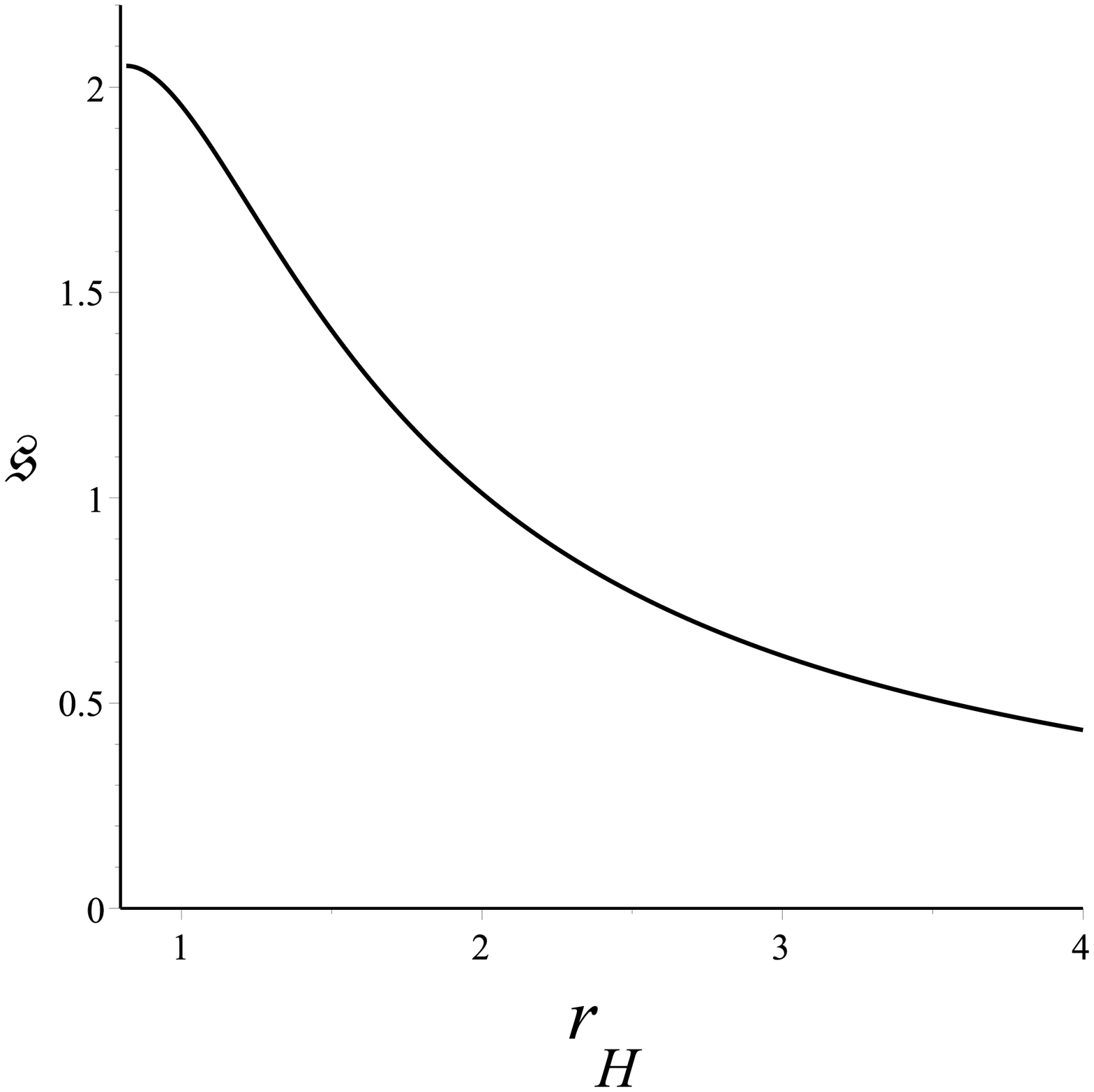}
\caption{Specific entropy of a ``large'' AdS$_5$-Reissner-Nordstr\"{o}m black hole with temperature approximately $1\%$ higher than the minimal possible temperature of an AdS$_5$-Schwarzschild black hole, using units in which $L = 1$. On the physical domain (to the right of the radius of the AdS$_5$-Schwarzschild black hole of that temperature) the function is decreasing. }
\end{figure}

In short, unless one specifically chooses the temperature at a one percent level of precision, the specific entropy of an asymptotically AdS$_5$-Reissner-Nordstr\"{o}m black hole of fixed temperature (above the minimum possible value for an AdS$_5$-Schwarzschild black hole) is always a decreasing function of the charge on the physical domain.

We see directly from equation (\ref{Q}) that the specific entropy can be made arbitrarily small by taking $r_{\textsf{H}},$ that is, the charge, sufficiently large. The physical meaning of this is as follows.

The parameter dual to the black hole charge is essentially the \emph{baryonic chemical potential}, $\mu_{\textsf{B}}$, which measures the disparity between particles and antiparticles; one can also think of it as a measure of the mass density of ultra-compact matter, as in neutron star cores. The relation between $\mathcal{Q}$ and $\mu_{\textsf{B}}$ is found by evaluating the asymptotic value of the electromagnetic one-form: see \cite{kn:clifford}. One finds that
\begin{equation}\label{QQ}
\mu_{\textsf{B}}\;=\;{k_5\mathcal{Q}\over 4\pi^2r_{\textsf{H}}^2}.
\end{equation}
Large values of the charge correspond to large values of both the numerator and the denominator here, if the temperature is fixed as usual. But we can clarify by combining this with equation (\ref{O}), obtaining
\begin{equation}\label{QQQ}
{32\pi \ell_5^3 \mu_{\textsf{B}}^2\over 3 k_5}\;=\;{4 r_{\textsf{H}}^2\over L^2}\;-\;4\pi T_{\textsf{AdSRN}_5} r_{\textsf{H}}\;+\;2.
\end{equation}
The larger root of the quadratic on the right side corresponds to the value of $r_{\textsf{H}}$ for a ``large'' AdS$_5$-Schwarzschild black hole; the quadratic increases indefinitely from there. That is, $\mu_{\textsf{B}}$ is a monotonically increasing function of the charge, and so large charge corresponds to large baryonic chemical potential. The holographic model predicts that matter at extremely high values of the baryonic chemical potential can have an arbitrarily small specific entropy compared to matter with smaller chemical potentials at the same temperature. It might be possible to explain this low specific entropy using the properties of the ``colour-flavour locked'' state thought to exist at the highest densities \cite{kn:alf}.

Throughout this discussion, we have assumed that we begin with an electrically neutral black hole, and then studied the consequences of gradually increasing the charge from zero. However, it is of course possible to begin with a black hole which is \emph{already} charged, and such a black hole can have a temperature below the AdS$_5$-Schwarzschild minimum $\sqrt(2)/(\pi L)$. In the dual system, this would correspond to temperatures such that the matter in question not strongly coupled at low temperatures and zero baryonic chemical potential, but it becomes strongly coupled (``quark matter'') at sufficiently high $\mu_{\textsf{B}}$. This is thought to be what does indeed happen in ultra-dense matter, possibly in the cores of neutron stars; see again \cite{kn:alf}.

Allowing lower temperatures than $\sqrt{2}/(\pi L)$ means that we can extend the range of permitted values for $r_{\textsf{H}}$ below those shown in Figure 4, and this might mean that there is a physically interesting range of electric charges for which the specific entropy is an \emph{increasing} function of the charge. However, there are two points to be borne in mind here.

The first is that, as we saw above, at any temperature, the specific entropy can only be an increasing function on the domain of $r_{\textsf{H}}$ values between zero and $\sqrt{2/3}\,L$. However, for such small values of $r_{\textsf{H}}$, equation (\ref{QQQ}) requires $\mu_{\textsf{B}}$ to be likewise small\footnote{An elementary argument shows that, for $0 < r_{\textsf{H}} < \sqrt{2/3}\,L$ and $0 < T_{\textsf{AdSRN}_5} < \sqrt{2}/(\pi L),$ the dimensionless quantity on the left in (\ref{QQQ}) is bounded above by $14/3$; in fact, for all but the smallest values of the temperature, it is considerably smaller than that.}. But, in reality, low-temperature quark matter only exists at extremely high values of the baryonic chemical potential. Admittedly, one would need a detailed analysis, with explicit data values, to define ``small'' in a precise way here, but it seems very unlikely that this domain of parameter values is anywhere near the physical domain.

Secondly, to show that this parameter domain is physical, one would need a detailed investigation of the stability of relatively cold AdS$_5$-Reissner-Nordstr\"{o}m black holes. In particular, one would need to understand the effects of electromagnetic superradiance \cite{kn:super}. On the basis of our findings in the analogous AdS$_5$-Kerr situation (see below) we suspect that the upshot would be that the conclusions of this section would not be greatly affected; that is, these unusual black holes are probably unstable. However, a more detailed analysis would be required to prove this.

In any case, we do not have an analogue of the Barnett effect in this case, so we have no reason to expect on holographic grounds that the specific entropy should decrease for all (physical) ranges of values of the baryonic chemical potential. The fact that, in nearly all cases, it actually does, is interesting; it suggests that the charged case might repay further investigation.

Here, however, our primary concern is with rotating black holes and vortical strongly coupled matter, to which we now turn.

\addtocounter{subsection}{1}
\section* {\large{\textsf{5.2. AdS$_5$-Kerr, Fixed Temperature}}}
The metric for the AdS$_5$-Kerr black hole (in the case of rotation around a single axis, the case with a clear holographic interpretation, therefore the only one we shall consider) takes the form \cite{kn:hawk,kn:cognola,kn:gibperry}
\begin{flalign}\label{R}
g(\textsf{AdSK}_5)\; = \; &- {\Delta_r \over \rho^2}\left[\,\m{d}t \; - \; {a \over \Xi}\,\m{sin}^2\theta \,\m{d}\phi\right]^2\;+\;{\rho^2 \over \Delta_r}\m{d}r^2\;+\;{\rho^2 \over \Delta_{\theta}}\m{d}\theta^2 \\ \notag \,\,\,\,&+\;{\m{sin}^2\theta \,\Delta_{\theta} \over \rho^2}\left[a\,\m{d}t \; - \;{r^2\,+\,a^2 \over \Xi}\,\m{d}\phi\right]^2 \;+\;r^2\cos^2\theta \,\m{d}\psi^2 ,
\end{flalign}
where
\begin{eqnarray}\label{S}
\rho^2& = & r^2\;+\;a^2\cos^2\theta, \nonumber\\
\Delta_r & = & \left(r^2+a^2\right)\left(1 + {r^2\over L^2}\right) - 2M,\nonumber\\
\Delta_{\theta}& = & 1 - {a^2\over L^2} \, \cos^2\theta, \nonumber\\
\Xi & = & 1 - {a^2\over L^2}.
\end{eqnarray}
Here $L$ is the background AdS curvature length scale, and $M$ and $a$ are purely geometric parameters (with units of length squared and length, respectively) which are \emph{not} equal or even simply related to the physical mass and the angular momentum per unit mass (see below). The angular coordinates on the (topological) $r = $ constant three-spheres are as before.

The quantity $\Xi$ plays an important role in the sequel, so let us explain its origin\footnote{$\Xi$ can be either strictly positive or strictly negative. The negative case ($a/L > 1$) is important \cite{kn:104}, but (assuming the validity of Cosmic Censorship) it cannot be attained by continuously spinning up the black hole. As we are interested in a holographic application to a system which \emph{is} obtained through a process of ``spinning up'' the QGP,  we only consider the strictly positive case, $a < L$.}.

Let us choose fixed values $t = t^*, \psi = 0, r = r^*$; then $\theta$ and $\phi$ can be interpreted as ordinary polar coordinates on a two-dimensional hemisphere with pole at $\theta = 0;$ we can think of the hemisphere as rotating about the axis through this pole, in the $\phi$ direction. Now on this hemisphere take a circle centred on $\theta = 0$, located at some fixed value of $\theta$.

To lowest order in $\theta$, $\Delta_{\theta} \approx \Xi,$ and, if $\rho^*$ denotes the value of $\rho$ when $r = r^*,$ $\rho^{*2} \approx r^{*2} + a^2,$ so if $\theta$ is small, the radius is given approximately by ${\sqrt{r^{*2} + a^2}\,\theta \over \sqrt{\Xi}}.$ The presence of $\Xi$ here begins a cascade which affects most aspects of the physics of these black holes.

The circumference of this circle is
\begin{equation}\label{SS}
C\;=\;{2\pi\over \Xi\,\rho^*}\sqrt{-\,a^2\Delta_r^*\sin^4 \theta \;+\;\left(r^{*2} + a^2\right)^2\Delta_{\theta}\sin^2 \theta},
\end{equation}
where $\Delta_r^*$ is $\Delta_r$ evaluated at $r^*$. Notice that the factor of $1/\Xi$ on the right is due to its explicit appearance in the metric tensor.

To lowest order in $\theta,$ $C = {2\pi \sqrt{r^{*2} + a^2}\,\theta \over \sqrt{\Xi}}.$ The point is that there is a danger that, as $\theta$ approaches zero, the ratio of the circumference of the circle to its radius will not approach $2\pi,$ meaning that the hemisphere (that is, \emph{every} hemisphere of this form, throughout the exterior spacetime) develops a conical singularity. We see that this is avoided here, but \emph{only} because $\m{d}\phi$ is divided by $\Xi$ in each of its appearances in the metric. (If these factors of $\Xi$ had not been included, the apical angle of the cones would have been $2\arcsin (\Xi)$.)

The consequence of this, however, is that the area of the event horizon inevitably acquires a factor of $1/\Xi$, and hence so does the entropy of the black hole:
\begin{equation}\label{T}
S_{\textsf{AdS}_5\textsf{K}}\; =\; {\pi^2\left(r_{\textsf{H}}^2 + a^2\right)r_{\textsf{H}}\over 2\ell_5^3\,\Xi}.
\end{equation}
But now, if the First Law is to hold, the presence of $\Xi$ in the entropy means that it must also be present in the physical mass and the angular momentum \cite{kn:gibperry}.

The upshot is that the physical mass $\mathcal{M}$ is related to the geometric parameter $M$ in a complex manner:
\begin{equation}\label{U}
\mathcal{M}\;=\;{\pi M \left(2 + \Xi\right)\over 4\,\ell_5^3\,\Xi^2},
\end{equation}
while the physical angular momentum is given by
\begin{equation}\label{V}
\mathcal{J}\;=\;{\pi M a\over 2\,\ell_5^3\,\Xi^2}.
\end{equation}
The specific angular momentum $\mathfrak{j}$ is therefore given by
\begin{equation}\label{W}
\mathfrak{j}\;=\;{2 a \over 2 + \Xi}\;=\;{2 a \over 3 - \left(a^2/L^2\right)}.
\end{equation}

There are two important observations to be made regarding (\ref{W}). First, (\ref{W}) implies that $\mathfrak{j}$ is a monotonically increasing function of $a$; this will be useful later.

Secondly, one shows easily that $\mathfrak{j}/L < 1$ if and only if $a/L < 1,$ as we are requiring here. (For all $a$ with $0 < a/L < 1,$ $\mathfrak{j}/L$ is smaller than $a/L$.) This means that the asymptotic AdS$_5$ curvature scale $L$ has a physical interpretation as \emph{the upper bound on the possible specific angular momenta} attainable by spinning up such a black hole (and therefore, by holography, on the possible specific angular momenta of the strongly coupled matter at infinity). Intuition suggests that this upper bound is actually imposed by the requirement that the vortical motion should never lead to superluminal speeds. Let us confirm this intuition.

Consider the equator of the hemisphere we discussed above; it is located at $\theta = \pi/2,$ and it defines (in the usual way) an equator on the corresponding hemisphere at conformal infinity. Now take a massive particle with angular momentum per unit mass equal to $\mathfrak{j}$ fixed on this equator: it belongs to the matter which is holographically dual to the bulk black hole (which by definition also has an angular momentum to mass ratio equal to $\mathfrak{j}.$) It is possible (though not straightforward) to show \cite{kn:93} that the velocity of this particle, relative to a distinguished observer at infinity with zero angular momentum, is given by
\begin{equation}\label{WW}
v_{\mathfrak{j}}\;=\;{\mathfrak{j}\over L}\,\sqrt{{1\over 1\;+\;{\mathfrak{j}^2\over L^2}\,\Xi}}.
\end{equation}
Here $\Xi$ should be regarded as a function of $\mathfrak{j}$ (see below).

Thus we can regard $v_{\mathfrak{j}}$ as a (surprisingly complicated, but monotonically increasing) function of $\mathfrak{j}.$ However, it is easy to see that it is bounded above by unity as $\mathfrak{j}$ tends to $L$. Clearly $\mathfrak{j}/L < 1$ is just the expression of causality for the matter at infinity which is dual to the black hole: a black hole with $\mathfrak{j}/L$ approximately equal to unity corresponds holographically to matter at infinity which is rotating at a speed close to that of light, and this is the meaning of the fact that $L$ imposes an upper bound on possible specific angular momenta. We will return to this observation later.

For later use, note first that from (\ref{W}) we have
\begin{equation}\label{X}
a \;=\;{L^2\over \mathfrak{j}}\left[-1\;+\;\sqrt{1 + {3\mathfrak{j}^2\over L^2}}\right].
\end{equation}
This allows us to express $\Xi$ in terms of $\mathfrak{j}:$
\begin{equation}\label{Y}
\Xi\;=\;{2L^2\over \mathfrak{j}^2}\left[\sqrt{1\,+\,{3\mathfrak{j}^2\over L^2}}\;-\;1\;-\;{\mathfrak{j}^2\over L^2}\right].
\end{equation}

The equation to be solved for the radial coordinate at the horizon, $r_{\textsf{H}},$ takes a deceptively simple form:
\begin{equation}\label{Z}
\left(r_{\textsf{H}}^2+a^2\right)\left(1 + {r_{\textsf{H}}^2\over L^2}\right)\; - \; 2\,M\;=\;0;
\end{equation}
but, in using this, we have to bear in mind that $a$ is \emph{not} the specific angular momentum, and $M$ is \emph{not} the physical mass, of the black hole.

The Hawking temperature of the AdS$_5$-Kerr black hole is given \cite{kn:gibperry} by
\begin{equation}\label{AA}
T_{\textsf{AdSK}_5}\;=\;{r_{\textsf{H}}\left(1 + {r_{\textsf{H}}^2\over L^2}\right)\over 2\pi \left(r_{\textsf{H}}^2 + a^2\right)} + {r_{\textsf{H}}\over 2\pi L^2},
\end{equation}

This equation has a very remarkable consequence: exactly extremal AdS$_5$-Kerr black holes actually \emph{do not exist} under the assumptions we are making here. For the only way to attain zero temperature is clearly if $r_{\textsf{H}} = 0$, which means that the ring singularity at $r = 0,\; \theta = \pi/2$ is visible at null conformal infinity, violating Cosmic Censorship. This is in fact very reasonable from a thermodynamic point of view: it means that near-extremal black holes of this sort, with extremely low temperatures, have extremely small values of $r_{\textsf{H}}$, that is, from equation (\ref{T}), very small entropies, in harmony with the Third Law of thermodynamics. (As we saw earlier, the analogous statements are by no means true of asymptotically flat black holes.)

Thus the temperature cannot vanish for these black holes. Nevertheless, we continue to use the familiar expression, ``near-extremal'' in the case where the temperature is small but not zero. (We can actually be quite precise about this: see below.)

There are in fact several other reasons to avoid the extremal and even the near-extremal cases here.

Firstly, AdS$_5$-Kerr black holes correspond holographically to strongly coupled matter at zero baryonic chemical potential, and such matter \emph{necessarily} has very high temperatures. From a holographic point of view, then, extremal and near-extremal AdS$_5$-Kerr black holes, which of course are ``cold'', are of limited physical interest.

Secondly, there are general reasons for suspecting \cite{kn:naresh,kn:horo} that exactly extremal spheroidal black holes are pathological in various ways, and also that near-extremal black holes are never stable (this is the Weak Gravity Conjecture \cite{kn:motl,kn:kats,kn:NAH}).

Finally, there is strong evidence that all extremal and sufficiently near-extremal AdS$_5$-Kerr black holes are actually \emph{classically} unstable due to superradiance \cite{kn:reall,kn:super}, and thus of little interest to us here. (We will see later that this is never a problem for ``large'' AdS$_5$-Kerr black holes.)

With all this preparation, we now proceed to our main goal, the analysis of the behaviour of the specific entropy of such a black hole.

The entropy of these black holes was given above, equation (\ref{T}), and so (using (\ref{U})) we find that the specific entropy is
\begin{equation}\label{BB}
\mathfrak{s}_{\textsf{AdSK}_5}\;=\;{2\pi r_{\textsf{H}}\left(r_{\textsf{H}}^2+a^2\right)\Xi\over M\left(2 + \Xi\right)}.
\end{equation}
Eliminating $M$ by means of (\ref{Z}), we have
\begin{equation}\label{CC}
\mathfrak{s}_{\textsf{AdSK}_5}\;=\;{4\pi r_{\textsf{H}}\Xi\over \left(2 + \Xi\right)\left(1 + {r_{\textsf{H}}^2\over L^2}\right)}.
\end{equation}

For fixed non-zero temperature, the equation in (\ref{AA}) is a cubic in $r_{\textsf{H}},$ and so it can be solved explicitly for $r_{\textsf{H}}$ in terms of $T_{\textsf{AdSK}_5}, a,$ and $L$. This can be substituted into (\ref{CC}), and equations (\ref{X}) and (\ref{Y}) can be used to eliminate $a$ and $\Xi$, and so we can express $\mathfrak{s}_{\textsf{AdSK}_5}$ explicitly as a function of $\mathfrak{j}$. An example is presented in Figure 6.

\begin{figure}[!h]
\centering
\includegraphics[width=0.6\textwidth]{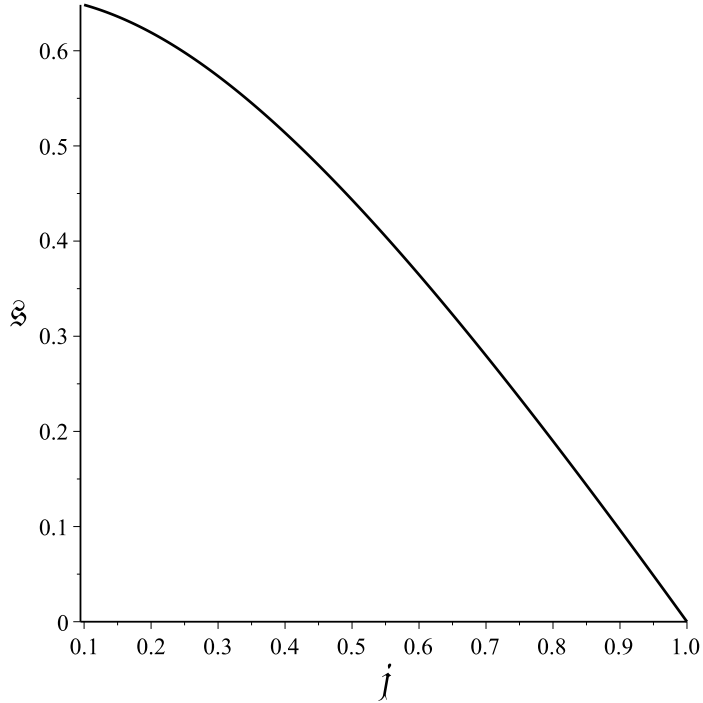}
\caption{Specific entropy of a five-dimensional asymptotically AdS$_5$ Kerr black hole with fixed temperature 2 in units with $L = 1,$ as a function of the specific angular momentum.}
\end{figure}

Clearly, in this instance, the specific entropy decreases with the specific angular momentum, as expected.

While the expression for $\mathfrak{s}_{\textsf{AdSK}_5}$ as a function of $\mathfrak{j}$ can be presented explicitly, it is of such length and complexity that it would be pointless to do so. (This extreme complexity is due to the presence of the factors of $\Xi,$ the necessity of which was explained earlier.) It is much more informative to proceed along the lines of our discussion of the AdS$_5$-Reissner-Nordstr\"{o}m geometry. That is, we use $r_{\textsf{H}}$ as a proxy for $\mathfrak{j}\,$; this is sufficient to determine whether the specific entropy decreases with increasing specific angular momentum. The procedure is as follows.

First, recall that $\mathfrak{j}$ and $a$ are monotonically increasing functions of each other: any increase in one faithfully represents an increase in the other. Thus we can take $a$ as our variable when discussing the effects of increasing the specific angular momentum.

Next, equation (\ref{AA}) allows us to regard $r_{\textsf{H}}$ as a simple function of $a$, when the temperature is fixed. As in the charged case, one finds that, provided that the temperature is at least $\sqrt{2}/(\pi L),$ there are two\footnote{Actually, there can be three, but the smallest of the three corresponds to a black hole which is never stable: see below.} possible values for $r_{\textsf{H}}$ for small values of $a$, corresponding as before to ``large'' and ``small'' AdS$_5$-Schwarzschild black holes. (For larger values of $a$, only the ``large'' branch exists.) A typical graph of $r_{\textsf{H}}$ as a function of $a$ is shown in Figure 7.

\begin{figure}[!h]
\centering
\includegraphics[width=0.6\textwidth]{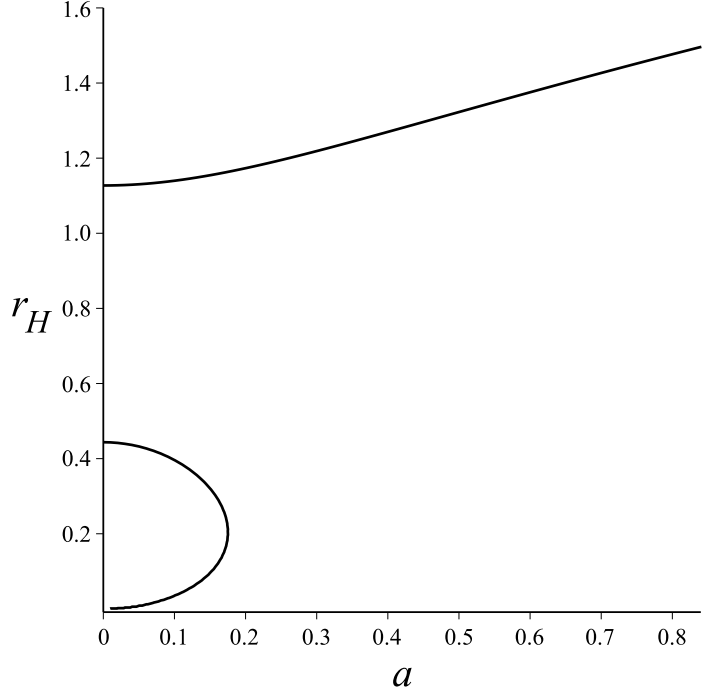}
\caption{Radial coordinate at horizon, with $L = 1, T_{\textsf{AdSK}_5} = 0.5$ . The domain is the physical one, corresponding to $0 \leq a < 1$, that is, to $0 \leq \mathfrak{j} < 1.$}
\end{figure}

We see that, if we take a ``large'' AdS$_5$-Schwarzschild black hole and begin to spin it up (that is, to increase $\mathfrak{j}$) at fixed temperature, this will increase $a$, and this in turn will increase $r_{\textsf{H}}$. Thus, as claimed, we can represent any increase in the specific angular momentum by an increase in $r_{\textsf{H}}$.

We can now confirm the absence of a superradiant instability here. Recall \cite{kn:reall} that superradiance is avoided if the angular velocity $\Omega$ of the outer horizon satisfies $\Omega L < 1.$ This condition is shown in Figure 8 as the dotted-dashed line: superradiance does not occur for parameter values corresponding to points above that line.

\begin{figure}[!h]
\centering
\includegraphics[width=0.6\textwidth]{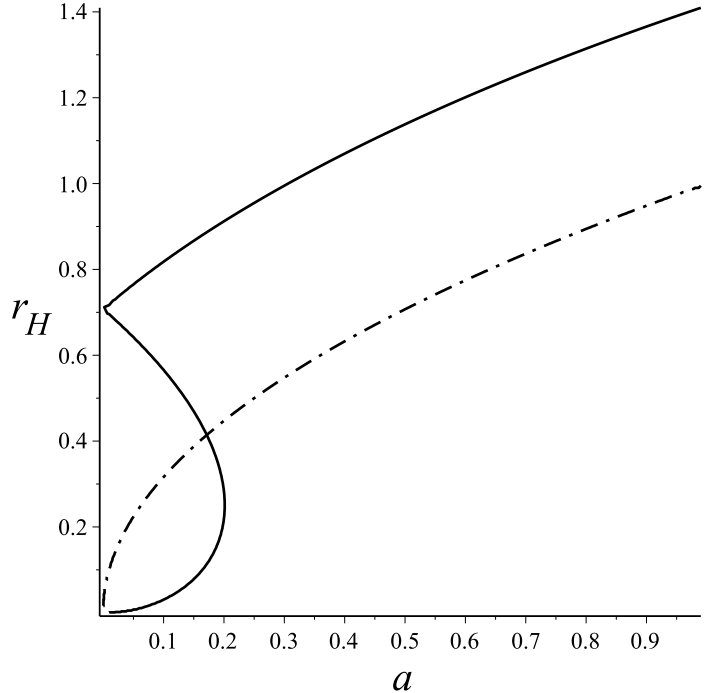}
\caption{AdS$_5$-Kerr black holes are stable against superradiance for all pairs $\left(a, r_{\textsf{H}}\right)$ lying above the dotted-dashed curve. The other curves show the relation between $a$ and $r_{\textsf{H}}$ for a black hole with temperature as small as possible in the non-rotating asymptotically AdS$_5$ case, with $0 \leq a < 1,$ corresponding to $0 \leq \mathfrak{j} < 1.$ $L$ has been set equal to unity throughout.}
\end{figure}

The solid lines show, as in Figure 7, $r_{\textsf{H}}$ as a function of $a$, but now in the case where the temperature is as low as possible for an AdS$_5$-Schwarzschild black hole. Clearly the branch describing ``large'' black holes is always above the dotted-dashed line, and therefore this is true when the temperature is above the minimum, since higher temperatures lift this branch higher. In short, all ``large'' AdS black holes remain stable against superradiance, for all values of the specific angular momentum and all temperatures for which they exist. (On the other hand, the smallest of the three values of $r_{\textsf{H}}$ which are possible when $a$ is sufficiently small is always ruled out by this criterion; some, but not all, ``small'' black holes are likewise excluded.)

Let us now examine the specific entropy, confining attention to the ``large'' case.

The range of values for $r_{\textsf{H}}$ is as follows. Of course, when $a = \mathfrak{j} = 0$ we obtain the value for a ``large'' AdS$_5$-Schwarzschild black hole, as given in equation (\ref{QMINUS}). When $a$ and $\mathfrak{j}$ are close to $L$ (their upper bound) then, from equation (\ref{AA}), we see by inspection that $r_{\textsf{H}}$ approaches $\pi T_{\textsf{AdSK}_5} L^2.$ This, then, is the physical range for $r_{\textsf{H}}$: ${L^2\over 2}\,\left(\pi T_{\textsf{AdSK}_5} + \sqrt{\pi^2 T_{\textsf{AdSK}_5}^2 - (2/L^2)}\right)$ to (nearly) $\pi T_{\textsf{AdSK}_5} L^2.$

The strategy now is to express the specific entropy, as given in equation (\ref{CC}), as a function of $r_{\textsf{H}}$ alone.

To do this, we solve (\ref{AA}) for $a$ as a function of $r_{\textsf{H}}$. The result, after some simplifications, is
\begin{equation}\label{DD}
a\;=\;{\frac {\sqrt {\left( 2\,\pi L^2 \,T_{\textsf{AdSK}_5}-r_{\textsf{H}} \right) r_{\textsf{H}} \left( -\,2\,{L}^
{2}\pi \,T_{\textsf{AdSK}_5}r_{\textsf{H}}+{L}^{2}+2\,{r_{\textsf{H}}}^{2} \right) }}{2\,\pi L^2 \,T_{\textsf{AdSK}_5}-r_{\textsf{H}}}}.
\end{equation}
Substituting this into (\ref{CC}), we find after simplification
\begin{equation}\label{EE}
\mathfrak{s}_{\textsf{AdSK}_5}\;=\;{\frac {4\pi r_{\textsf{H}} \, \left( \pi L^2 \,T_{\textsf{AdSK}_5}-r_{\textsf{H}} \right) {L}^{2}}{3\,\pi L^4 \,T_{\textsf{AdSK}_5}+\pi L^2 \,T_{\textsf{AdSK}_5}{r_{\textsf{H}}}^{2}-2\,{L}^{2}r_{\textsf{H}}-{r_{\textsf{H}}}^{3}}};
\end{equation}
this is vastly simpler than the explicit expression for the specific entropy in terms of $\mathfrak{j}$. We just have to bear in mind that ``increasing $\mathfrak{j}$ from $0$ towards its upper bound, $L$'' corresponds to ``increasing $r_{\textsf{H}}$ from ${L^2\over 2}\,\left(\pi T_{\textsf{AdSK}_5} + \sqrt{\pi^2 T_{\textsf{AdSK}_5}^2 - (2/L^2)}\right)$ towards $\pi T_{\textsf{AdSK}_5} L^2.$'' (Note that the cubic in the denominator in (\ref{EE}) has only one real root, and one can show that that root always lies outside the permitted range for $r_{\textsf{H}}$.)

Let us begin with the lowest possible temperature in the non-rotating case, $\sqrt{2}/\pi \approx 0.45$ in units with $L = 1.$ The graph of $\mathfrak{s}_{\textsf{AdSK}_5}$ is shown in Figure 9.

\begin{figure}[!h]
\centering
\includegraphics[width=0.6\textwidth]{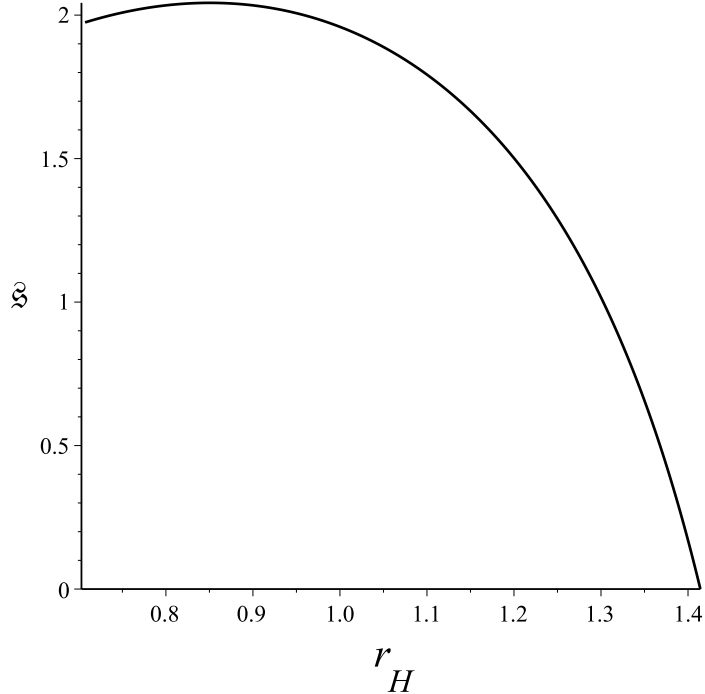}
\caption{Specific entropy of an AdS$_5$-Kerr black hole with fixed temperature equal to the smallest possible AdS$_5$-Schwarzschild temperature $\sqrt{2}/\pi \approx 0.450$ in units with $L = 1,$ as a function of $r_{\textsf{H}}$. The domain is the physical one, $\approx 0.707 \leq r_{\textsf{H}} <  \approx 1.414,$ corresponding to $0 \leq \mathfrak{j} < 1.$}
\end{figure}

As in the AdS$_5$-Reissner-Nordstr\"{o}m case, we find that it is possible for the specific entropy to increase slightly if the specific angular momentum increases from zero to a small value; but then it decreases for all larger values. However, as before, we find that even a slight increase in the temperature eliminates this odd behaviour. For example, if the temperature is increased by just one percent to $\approx 0.455,$ the physical domain is such that $\mathfrak{s}_{\textsf{AdSK}_5}$ is a decreasing function throughout it: see Figure 10.

\begin{figure}[!h]
\centering
\includegraphics[width=0.6\textwidth]{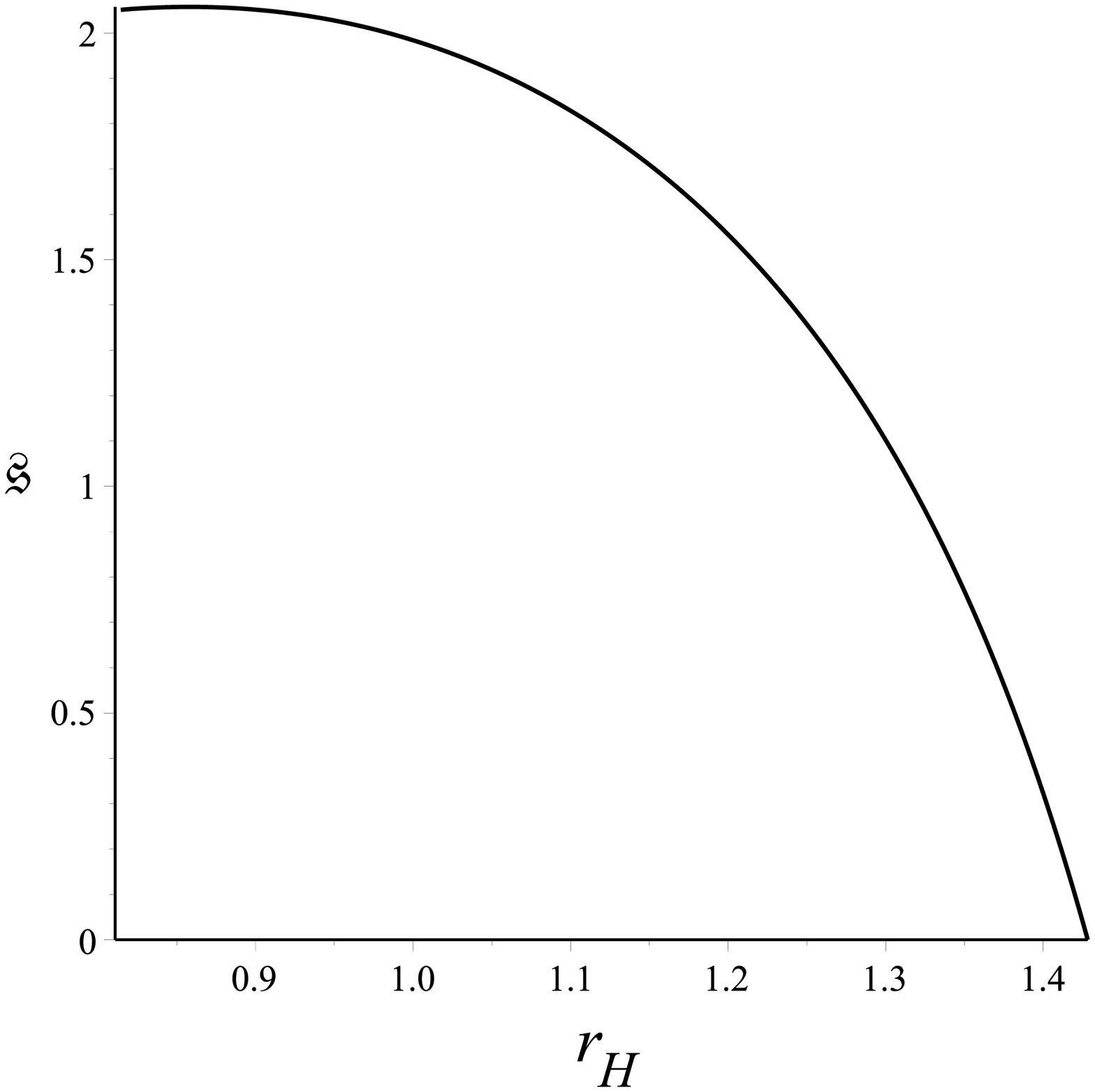}
\caption{Specific entropy of an AdS$_5$-Kerr black hole with fixed temperature $\approx 0.455$ in units with $L = 1,$ as a function of $r_{\textsf{H}}$. The domain is the physical one,  $\approx 0.814 \leq r_{\textsf{H}} <  \approx 1.428,$ corresponding to $0 \leq \mathfrak{j} < 1.$}
\end{figure}

We conclude, then, that the specific entropy of these black holes decreases, at fixed temperature, as the specific angular momentum increases from zero, \emph{unless} the temperature is just slightly (less than 1$\%$) greater than its minimum possible value for a non-rotating AdS$_5$ black hole. In a holographic picture of quark matter, this translates to the statement that a vortical QGP (nearly) always has a lower specific entropy than the QGP produced by central collisions in the same beam, in agreement with expectations based on the quark matter version of the Barnett effect.

The exceptional cases with temperature less than $1\%$ above the AdS$_5$-Schwarzschild minimum are of more mathematical than physical interest. For it is well known that, in the regime of small baryonic chemical potential, the transition to the QGP is not a sharp phase change: it is a continuous ``crossover'' \cite{kn:franc}. That is, the transition is not very precisely defined, so the phenomenon of increasing specific entropy for these very special black holes has no meaningful counterpart in the real QGP. The exceptional cases are however theoretically interesting in that their existence shows that the \emph{generic} decrease of the specific entropy with increasing specific angular momentum does not have a straightforward explanation in terms of basic black hole thermodynamics.

As in the preceding Section, we have, throughout this discussion, only considered temperatures which are possible for non-rotating black holes. This is justified in the present case, because in the dual theory we want to compare strongly coupled matter produced in central heavy-ion collisions (with negligible angular momentum density) with its counterparts produced in peripheral collisions. However, as before, it is instructive to consider the case where the dual matter is ``already'' rotating. This means that we allow temperatures below\footnote{This means that we are considering collisions at such low impact energies that a quark-gluon plasma is not formed in central collisions, but \emph{is} formed in some peripheral collisions. No such effect has been observed, but presumably this is possible.} $\sqrt{2}/(\pi L).$

At these temperatures, the two parts of the $r_{\textsf{H}}$ vs. $a$ curve merge, and the curve pulls away from the vertical axis, so that there is no longer a distinction between ``large'' and ``small'' black holes; if it is $1/(\pi L)$ or lower, then the entire merged curve lies below the dotted-dashed line indicating superradiant instability. In this case (which includes extremal AdS$_5$-Kerr black holes, long known \cite{kn:reall} to be unstable against superradiance) the black holes are unstable for \emph{all} non-zero values of the specific angular momentum. Thus, a temperature below $1/(\pi L)$ is a good \emph{definition} of ``near-extremal'' for these black holes: then we can say that all ``near-extremal'' AdS$_5$-Kerr black holes are classically unstable under all circumstances. We do not consider this case further.

When the temperature lies between $1/(\pi L)$ and $\sqrt{2}/(\pi L),$ $r_{\textsf{H}}$ is apparently allowed to go down to values such that the specific entropy is an increasing function of the specific angular momentum. In practice, however, that is not the case, at least not to any significant extent. For the black hole can be stable only if the curve describing $r_{\textsf{H}}$ as a function of $a$ lies above the curve below which superradiance occurs. However, that only happens for a narrow range of $r_{\textsf{H}}$ values. For example, in Figure 11, which portrays the case where the temperature is $80\%$ of the minimal AdS$_5$-Schwarzschild value, the black hole is only stable when $r_{\textsf{H}}$ lies in the range from $\approx 0.602$ to $\approx 1.131$. As before, then, the domain in which the specific entropy is an increasing function is almost completely excluded, as one can see from Figure 12; so in fact this case does not differ very greatly from the case of higher temperatures.

\begin{figure}[!h]
\centering
\includegraphics[width=0.6\textwidth]{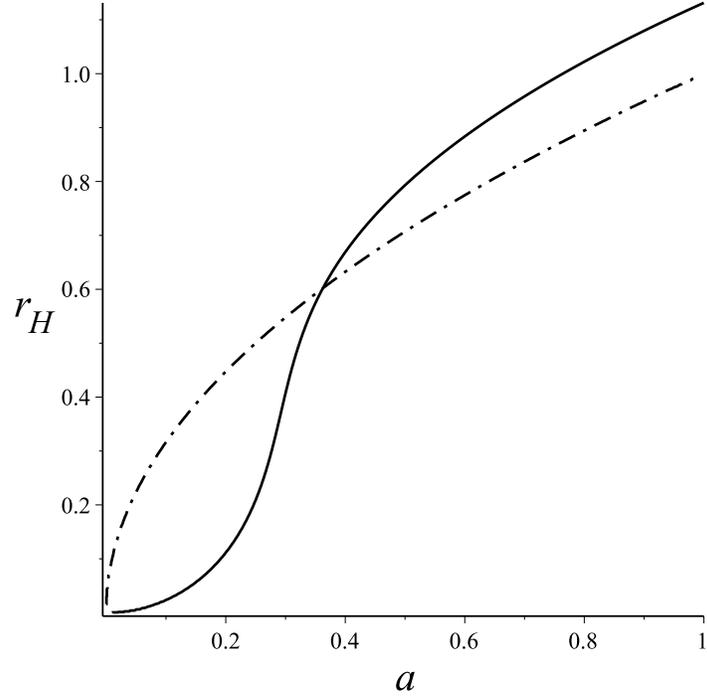}
\caption{The relation between $r_{\textsf{H}}$ and $a$ when the temperature is $80\%$ of the minimal AdS$_5$-Schwarzschild value, shown, as in Figure 8, with the line below which the black hole is unstable to a superradiant instability. Units are such that $L = 1.$}
\end{figure}

\begin{figure}[!h]
\centering
\includegraphics[width=0.6\textwidth]{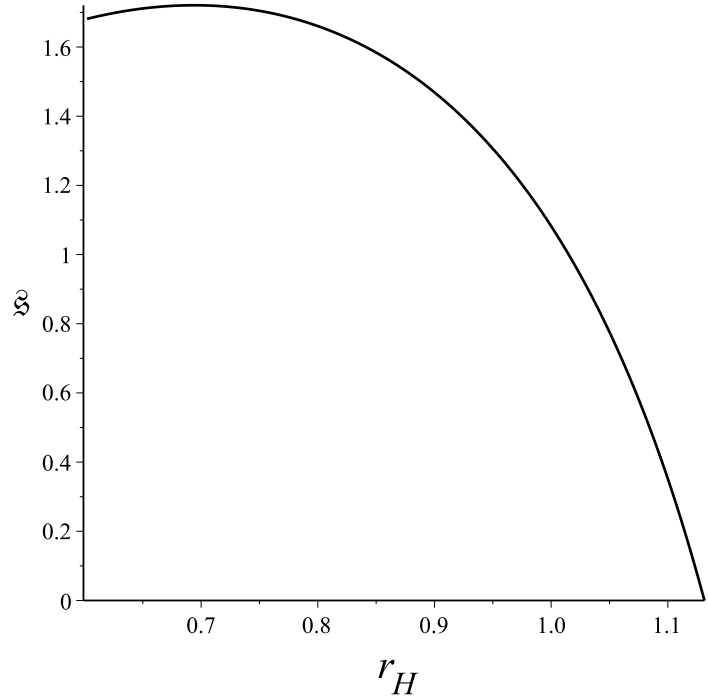}
\caption{The specific entropy of the black hole, with parameters as in Figure 11. The domain is the one on which the black hole is stable against superradiance.}
\end{figure}

\addtocounter{section}{1}
\section* {\large{\textsf{6. Conclusion}}}
This work is motivated by the simple suggestion that, if we wish to identify black hole microstates, then we might do well to study the statistical mechanics of some similar system where the microphysics is better understood. In the case of rotating black holes, the natural choice is the ``vortical QGP'' which has been intensively studied experimentally \cite{kn:STARcoll}.

In that case, it is possible to compare vortical strongly coupled matter with non-vortical strongly coupled matter at the same temperature by simply comparing central and peripheral collisions in the same beam of colliding heavy ions. Here a very simple argument based on the physics underlying the Barnett effect leads us to expect a decreasing specific entropy with increasing specific angular momentum. We found that the holographically related system, an AdS$_5$-Kerr black hole, behaves generically in precisely that manner.

Over-simplifying somewhat: the answer to our question, ``why does rotation reduce (AdS) black hole specific entropy?'' is, ``because rotation has that effect on the dual strongly coupled matter, for a reason which in that case is quite clear.''

Whether one should regard this as a \emph{satisfactory} answer is perhaps a philosophical question, which we may leave to one side. Our point is that (in this particular case) holography provides a route towards a better understanding, at least of black hole entropy and possibly ultimately of black hole microstates.

We conclude with the following rather suggestive observation.

In the asymptotically flat cases (whether mass or temperature was fixed) we always found that increasing the specific angular momentum, or the charge, reduces the specific entropy of the black hole. But there was always a limit to the extent of the reduction: for example, in the four-dimensional fixed-temperature asymptotically flat Kerr case, we found that, no matter what value was chosen for the specific angular momentum, the specific entropy could never fall below about 62$\%$ of the value for a Schwarzschild black hole of that temperature. (See Figure 2.)

In the AdS$_5$-Kerr case, however, there is no such restriction: one sees from equation (\ref{EE}) that the specific entropy can be reduced to any prescribed positive value, however small, by choosing $r_{\textsf{H}}$ sufficiently close to $\pi T_{\textsf{AdSK}_5} L^2;$ that is, by choosing $\mathfrak{j}$ sufficiently close to its upper bound, $L$. (This is clear in Figures 6, 9, 10, and 12.) This is true at \emph{any} fixed temperature.

This is reminiscent of the situation in quantum statistical mechanics where the energy spacing of the ground state and the first excited state is greater than the temperature: the entropy can be very small even if the temperature is not. Thus it seems that very extreme specific angular momenta can give rise to this situation for rotating AdS black holes. That is, at any given temperature, it is possible, by increasing the specific angular momentum suitably, to drive up the spacing between the black hole ground state and its ``first excited state'' to a value beyond that temperature, even if the latter is large. Further study of this phenomenon might lead to an identification of this ``excited state'' and perhaps then to an insight into the nature of black hole microstates.

Unfortunately, heavy ion collisions do not offer an observational guide to this regime of values for $\mathfrak{j}.$ To see why, recall that, in order for $\mathfrak{j}$ to be close to $L$, the rotational speed of the dual matter at conformal infinity would have to be close to that of light (equation (\ref{WW})); presumably this would correspond, in the actual vortical QGP, to motion close to the speed of light in the vortices. In the experiments reported in \cite{kn:STARcoll}, however, that is not the case. A typical angular velocity for vortices in those experiments is around $0.03 \;\m{fm}^{-1}$. This is indeed a gigantic vorticity (about $9\,\times 10^{21}\,\cdot\,$s$^{-1}$) by normal standards, but, for these systems, with a radius of at most a few femtometres, this does \emph{not} imply velocities close to that of light \cite{kn:declan}; and so $L$ must still be well above the specific angular momenta in these experiments.

One might propose to investigate the situation in collisions at higher impact energies, such as those studied by the ALICE collaboration at the LHC; but, for reasons which are only partly understood \cite{kn:jiang}, the polarizations of $\Lambda$/$\overline{\Lambda}$ hyperons actually \emph{decrease} with increasing impact energy, and in fact they are quite unobservable at the ALICE energies \cite{kn:bed}. If this changes with further observations, it would be very interesting to try to determine whether such high-temperature systems do in fact have very low specific entropies when the specific angular momenta are extremely large.

\addtocounter{section}{1}
\section*{\large{\textsf{Acknowledgements}}}
The author is grateful to Dr. Soon Wanmei for useful discussions.

\end{document}